# Soft chemical control of superconductivity in lithium iron selenide hydroxides $Li_{1-x}Fe_x(OH)Fe_{1-y}Se$.

Hualei Sun,[a,b] Daniel N. Woodruff,[a] Simon J. Cassidy,[a,c] Genevieve M. Allcroft,[a] Stefan J. Sedlmaier,[a] Amber L. Thompson,[a] Paul A. Bingham,[d] Susan D. Forder[d] Simon Cartenet,[d] Nicolas Mary,[d] Silvia Ramos,[e] Francesca R. Foronda,[f] Benjamin H. Williams,[f] Xiaodong Li,[b] Stephen J. Blundell[f] and Simon J. Clarke*[a]

[a] Department of Chemistry, University of Oxford, Inorganic Chemistry Laboratory, South Parks Road, Oxford, OX1 3QR, UK. [b]Beijing Synchrotron Radiation Facility, Institute of High Energy Physics, Chinese Academy of Science, Beijing 100049, China. [c]Diamond Light Source Ltd., Harwell Science and Innovation Campus, Didcot, OX11 0DE, UK. [d]Materials and Engineering Research Institute, Faculty of Arts, Computing, Engineering and Sciences, Sheffield Hallam University, City Campus, Howard Street, Sheffield, S1 1WB,UK. [e]School of Physical Sciences, Ingram Building, University of Kent, Canterbury, Kent, CT2 7NH, UK. [f]Department of Physics, University of Oxford, Clarendon Laboratory, Parks Road, Oxford, OX1 3PU, UK.

**ABSTRACT:** Hydrothermal synthesis is described of layered lithium iron selenide hydroxides $Li_{1-x}Fe_x(OH)Fe_{1-y}Se$ ($x \sim 0.2$; $0.02 < y < 0.15$) with a wide range of iron site vacancy concentrations in the iron selenide layers. This iron vacancy concentration is revealed as the only significant compositional variable and as the key parameter controlling the crystal structure and the electronic properties. Single crystal X-ray diffraction, neutron powder diffraction and X-ray absorption spectroscopy measurements are used to demonstrate that superconductivity at temperatures as high as 40 K is observed in the hydrothermally synthesised samples when the iron vacancy concentration is low ($y < 0.05$) and when the iron oxidation state is reduced slightly below +2, while samples with a higher vacancy concentration and a correspondingly higher iron oxidation state are not superconducting. The importance of combining a low iron oxidation state with a low vacancy concentration in the iron selenide layers is emphasised by the demonstration that reductive post-synthetic lithiation of the samples turns on superconductivity with critical temperatures exceeding 40 K by displacing iron atoms from the $Li_{1-x}Fe_x(OH)$ reservoir layer to fill vacancies in the selenide layer.

## Introduction.

Iron-based arsenide[1] and selenide superconductors are compounds where chemical control of the properties by isovalent or aliovalent substitution[1-4] reveals competing itinerant antiferromagnetic and unconventional superconducting states.[5,6] The almost-stoichiometric tetragonal polymorph of iron selenide, $Fe_{1.01}Se$, is a superconductor with a superconducting transition temperature $T_c$ of 8.5K.[7,8] Some FeSe derivatives exhibit higher $T_c$s[9] but often contain ordered arrays of iron site vacancies,[10,11] with superconductivity in minority regions.[12-14] In order to decrease the concentration of iron site vacancies in the FeSe layers, stoichiometric, superconducting FeSe itself has been used in the synthesis, at ambient temperatures and below, of intercalates using solutions of electropositive metals in ammonia.[15] These intercalates, which often superconduct at temperatures as high as 45 K contain variable electropositive metal and ammonia and amide contents and are the subject of current investigation.[16-19]

Recently layered lithium iron selenide hydroxides have been reported with $T_c$s of up to about 40 K.[20-22] Here we reveal the phase width in these hydrothermally synthesised compounds $Li_{1-x}Fe_x(OH)Fe_{1-y}Se$ ($x\sim0.2$; $0.02 < y < 0.15$) and control their compositions. We quantify the correlations between superconductivity and the concentration of iron vacancies in the selenide layer and the electron count of iron. We underline this by demonstrating that post-synthetic reductive lithiation displaces iron ions from the hydroxide layer "reservoir" into the selenide layer to reduce the iron deficiency in the selenide layers to zero and turn on bulk superconductivity with $T_c > 40$ K.

## Experimental Methods

*Synthesis.* The hydrothermal synthesis was adapted from that in ref. 20. Our approach differs from that previously reported in that we used tetragonal FeSe as the source of all the Se and most of the Fe in the synthesis. The Pourbaix diagram for iron and selenium is known from investigations of the contamination of natural waters[23] and reveals that under reducing conditions and at high *p*H values the formation of $H_2Se$ is suppressed and FeSe is stable. Accordingly the samples were synthesised under mildly reducing and extremely basic hydrothermal conditions obtained by incorporating high purity elemental iron into the syntheses along with FeSe, using a large excess of lithium hydroxide, and by excluding oxygen from the synthesis. Typically 6 mmol (0.8 g of tetragonal FeSe (synthesised from the elements (Fe ALFA 99.998 %; Se ALFA 99.999%) as described previously[8]), 140 mmol (6 g) of $LiOH\cdot H_2O$ (Aldrich 98%) and 5 ml of deionised and de-oxygenated water were loaded into a Teflon-lined steel autoclave of 18 $cm^3$ capacity together with variable amounts of additional iron powder. The autoclaves were tightly sealed and placed in a chamber furnace. The furnace was heated to 200 °C at 1°C per minute and the temperature was maintained for 12 days. The furnace was then turned off and allowed to cool naturally and the autoclaves were removed at room temperature. The autoclaves were opened in an argon-filled glove bag and the products were loaded into Schlenk tubes and washed three times with deionised and de-oxygenated water to remove soluble side products. Magnetic impurities were removed from some syntheses using a strong magnet. The samples were dried under vacuum and removed to an argon-filled glovebox. The synthesis was scalable in the 18 $cm^3$ autoclaves to produce 10 g of product by increasing the amount of FeSe and Fe in the synthesis 12 fold, increasing the amount of water to 7 $cm^3$ and maintaining the amount of $LiOH\cdot H_2O$ which remains in a large excess. Some of these samples were subsequently subjected to lithiation in which the powders were stirred in solutions of lithium in liquid ammonia at –30°C using a Schlenk line, with subsequent evaporation of the solvent and evacuation to yield the dried product which contained some $LiNH_2$ arising from decomposition of the $Li/NH_3$ solution. (*Caution:* ammonia has a vapour pressure of ~8 bar at ambient temperature and is highly toxic and flammable. The reactions with metal/ammonia solutions were performed in a fume hood. Pressure relief, via a mercury bubbler, for evaporating ammonia and any hydrogen formed in the reactions was always available.)

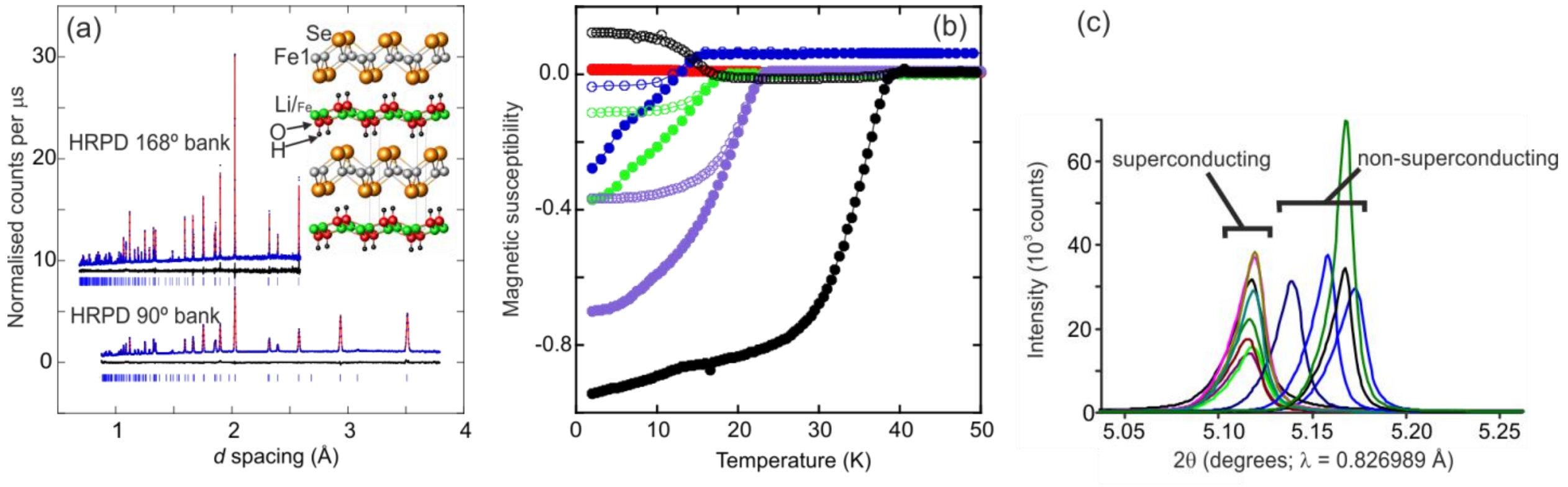

Figure 1. (a) The structure of the lithium iron selenide hydroxides and a typical refinement against neutron diffraction data (HRPD at ISIS) showing data (blue dots), calculated (red line), difference (black line) and reflection positions. Data for the 168° bank have been displaced by 9 units along the vertical axis. $Li_{1-x}Fe_x(OH)Fe_{1-y}Se$: Space group *P*4/*nmm* (No. 129) *a* ~ 3.8 A, *c* ~ 9.2 Å. Atomic positions (origin choice 2: inversion centre at origin): Fe: site 2*a* (¼, ¾, 0); Se: site 2*c* (¾, ¾, z~0.16); O: site 2*c* (¼, ¼, *z*~0.43); $Li_{1-x}Fe_x$: site 2*b* (¾, ¼, ½); H: site 2*c* (¼, ¼, z~0.33) (see Tables S1 & S2, Figures S1 and S2 and the crystallographic information file included in the ESI). (b) Magnetic susceptibility measurements showing the full range of behaviour spanned by hydrothermally synthesised samples. Zero-field-cooled (filled symbols) and field-cooled (open symbols) data are shown. Some samples showed a high normal state background due to minuscule amounts of magnetic impurities. The sample with the highest $T_c$ exhibits diamagnetism in the field-cooled measurement, but also shows a low temperature transition below 20 K which is presumed to arise from a magnetic impurity present at levels below the detection limit of our diffraction experiments. SI conventions were used in determining the dimensionless magnetic susceptibility. (c) Plot of the 001 reflection measured for a range of samples on I11 showing the correlation between the *c* lattice parameter and whether the compounds are superconducting. The values of the basal lattice parameter *a* spanned a range of 0.9 % and the values of the lattice parameter *c* spanned a range of 1.1 %. Small values of *a* corresponded to large values of *c*, so the unit cell volumes spanned just 0.8 %.

*Diffraction Measurements.* X-ray Powder Diffraction (XRPD) measurements used beam line I11 at the Diamond Light Source, UK with 0.8 Å X-rays and the multi analyser crystal detector bank. Neutron Powder Diffraction (NPD) measurements used the GEM and HRPD diffractometers at the ISIS Facility, UK. Single Crystal X-ray diffraction (SCXRD) was carried out on small (~10 × 10 × 1 μm) crystals (Figure S6)) using beam line I19 at Diamond using 0.68890 Å X-rays. *Ab initio* structure solution from SCXRD data was performed using SuperFlip[24] implemented within CRYSTALS,[25] with refinements performed using CRYSTALS. Refinements against powder diffraction data (Table S1; Figure 1, Figures S2 – S3)) were conducted using TOPAS Academic.[26]

*Magnetometry*. Measurements used Quantum Design MPMS SQUID magnetometers and measuring fields of 20 – 50 Oe to characterise the superconducting state and up to 7 T to probe the normal state susceptibilities. Samples were sequestered from air in gelatine capsules. Susceptibilities were corrected for the effect of demagnetising fields arising from the shape of the sample.[27]

*X-ray Absorption Spectroscopy*. Measurements were conducted in transmission mode on beamline B18 at Diamond with the samples sequestered from air and diluted with cellulose powder. All spectra were calibrated against an iron foil. The data were analysed using Athena and Artemis, part of the Demeter software package.[28]

*Muon-spin rotation spectroscopy*. 300 mg of powder was contained in a silver foil packet and was sequestered from air prior to loading into the helium atmosphere of the cryostat. Variable temperature measurements were carried out in applied transverse magnetic fields of up to 30 mT on the MuSR beamline at the ISIS facility.

## Results and discussion

**Hydrothermally synthesised samples.** The products of the hydrothermal reactions were black with metallic lustre and were examined with no further synthetic treatment. SQUID magnetometry (Figure 1(b)) carried out on samples from iron-rich syntheses (overall ratio of Fe:Se in the synthesis of 1.16:1; i.e. 1 mmol additional Fe for 6 mmol FeSe in the autoclave) revealed superconductivity with $T_c$s in the range 10–39K and variable shielding fractions. The use of smaller amounts of additional Fe (0 or 0.5 mmol Fe per 6 mmol of FeSe) produced non-superconducting products. The use of larger amounts of additional Fe led to significant contamination by iron oxide side products. The products were highly crystalline and appeared single phase using high resolution X-ray and neutron powder diffraction (Figure 1). Diffractograms were indexed on tetragonal cells in space group $P4/nmm$ with lattice parameters of $a$~3.8 Å and $c$~9.2 Å. Lattice parameters were found to be highly sample dependent and correlated with the occurrence, or not, of superconductivity (Figure 1(c), Tables S1–S2): superconductors from iron-rich syntheses had unit cell volumes < 133.2 Å$^3$ and $c/a$ > 2.43, while non-superconductors from iron-poor syntheses had cell volumes >133.2 Å$^3$ and $c/a$ < 2.43. *Ab initio* structure solution from SCXRD data yielded the chemically unsatisfactory structural model of ref 20 with iron selenide layers separated by "spacer" layers with a similar topology and with the atoms in this "spacer" layer (O and $Li/_{Fe}$ in Figure 1(a)) all appearing isoelectronic with oxygen. The shortest distances between the selenide ions and the nearest atoms (labelled O in Fig. 1(a)) in the "spacer" layer were 3.62 Å, only marginally shorter than the interlayer Se···Se distances of 3.71 Å in tetragonal FeSe[8] and marginally longer than the between-layer Se···Se distance of 3.58 Å in $TiSe_2$,[29] and thus longer than one would expect for Se···O non-bonded distances.

NPD data collected on bulk samples enabled a chemically sensible model to be obtained. The sites labelled $Li/_{Fe}$ in the "spacer" layers were approximately null scattering, and an additional region with a negative scattering density, corresponding to a hydrogen nucleus with full occupancy within the experimental uncertainty, was located about 1Å from the atoms in the "spacer" layers labelled as O in Figure 1(a).

Two samples were measured on the GEM neutron diffractometer at room temperature and 50 K. The refinements against data gathered at the two temperatures produced similar site occupancies showing that the wide $d$-spacing range available on the time-of-flight diffractometer, the high crystallinity, and the almost flat neutron form factor minimise parameter correlations in the refinements against these highly crystalline samples, and that this method is robust for determining site occupancy factors with a high precision. Single crystals extracted from several of the samples measured by NPD at room temperature were found to faithfully represent the bulk of the sample probed in the NPD experiments and the results of the refinements against all of our I19 SCXRD datasets are therefore also included in the analysis (Table S2).

For our wide range of different samples NPD and SCXRD together produced an unambiguous structural model with lithium/iron hydroxide layers containing a 0.8:0.2 Li:Fe disordered mixture (approximately null scattering for neutrons ($b_{Li}$ = –1.90 fm; $b_{Fe}$ = 9.45 fm)[30] and with an average electron count similar to that of oxygen) separating iron selenide layers (Figure 1(a)). While we were performing this work this conclusion was reported by other groups, each from analysis of a single composition.[21,22] Using our synthetic method, we obtained refined compositions $Li_{1-x}Fe_x(OH)Fe_{1-y}Se$ with $x$~0.2, and almost sample invariant (Figure 2(a)), and $y$ representing a 2–15% deficiency on the Fe1 site in the iron selenide layers. The Fe1 deficiency was similar within the uncertainty when measured using both NPD and SCXRD measurements on several sample batches spanning the range of lattice parameters, which suggests that in the samples described here it is a true deficiency and not the result of Li and Fe also sharing a vacancy-free site in the selenide layers as has been proposed in the analysis, by single crystal X-ray diffraction, of a single related composition examined elsewhere.[22] Further evidence that the iron site in the selenide layers in our samples carries a deficiency comes from the results of post-synthetic lithiation described below. The H-contents are similar for all samples within the uncertainty and the Se–H distances of about 3.1 Å between the selenide and hydroxide layers correspond well to those found for weak hydrogen bonding interactions.[17] Thermogravimetric analysis under dry $N_2$ was consistent with dehydration of $Li_{1-x}Fe_x(OH)Fe_{1-y}Se$ commencing at about 350°C (Figure S7).

The basal lattice parameter $a$ varies linearly with the occupancy of the Fe1 site in the selenide layers which ranges from 0.85(1) to 0.98(1) for the hydrothermally synthesised samples (Figure 2(a)). This Fe1 occupancy is the only significant compositional and structural difference between samples. Increasing the site occupancy strengthens the Fe–Fe bonding within the FeSe layer, shortening the lattice parameter. Key structural parameters for iron-based superconductors are the Fe−Fe distance in the plane (=$a/\sqrt{2}$), the Fe−$E$ ($E$ = chalcogen or pnictogen) bond length and the $E$−Fe−$E$ angles in the $FeE_4$ tetrahedra. For the current compounds the $FeSe_4$ tetrahedra are extremely squashed in the basal plane relative to the more regular tetrahedra found in iron arsenide superconductors,[31] as in FeSe[8] and its intercalates.[17] The Fe−Se distance is rather invariant across the series, and the change in $a$ lattice parameter is manifested in the Se−Fe−Se angles (Figure 2(b)).

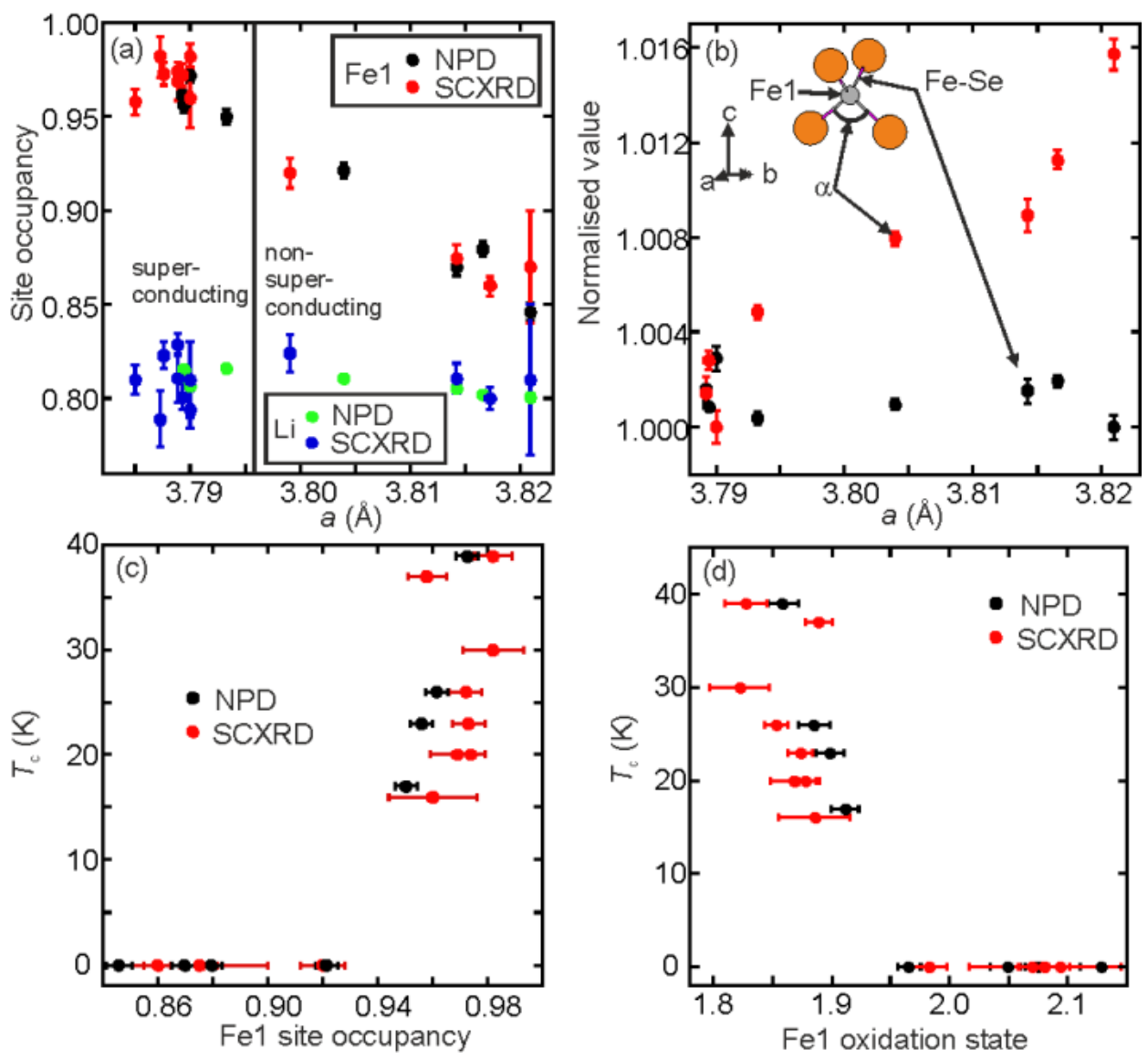

Figure 2. (a) Plot of basal lattice parameter, $a$, against site occupancy of Fe in the iron selenide layer and of Li in the hydroxide layer obtained from refinements against NPD and SCXRD data. Lattice parameters for the SCXRD samples were obtained at ambient temperature using synchrotron XRPD. The single crystal of the sample with the largest $a$ lattice parameter had an unusually large mosaic spread which is the likely origin of the relatively large errorbars on the refined occupancies. (b) Variation with $a$ (=$\sqrt{2}\times$Fe−Fe) of Fe−Se bond lengths (●) and the Se−Fe−Se angle of multiplicity two (●) (often denoted α) normalised against the smallest value in each series, obtained from NPD at ambient temperatures. The shape of the $FeSe_4$ tetrahedra is similar to that in other iron selenide superconductors and is characterised by being much more squashed in the basal plane than in iron arsenide superconductors (α has a value of about 103°). (c) The correlation of superconducting $T_c$ with refined Fe site occupancy in the selenide layers obtained from refinements against NPD and SCXRD. (d) Correlation of superconducting $T_c$ with Fe oxidation state obtained from the compositions refined from NPD and SCXRD measurements.

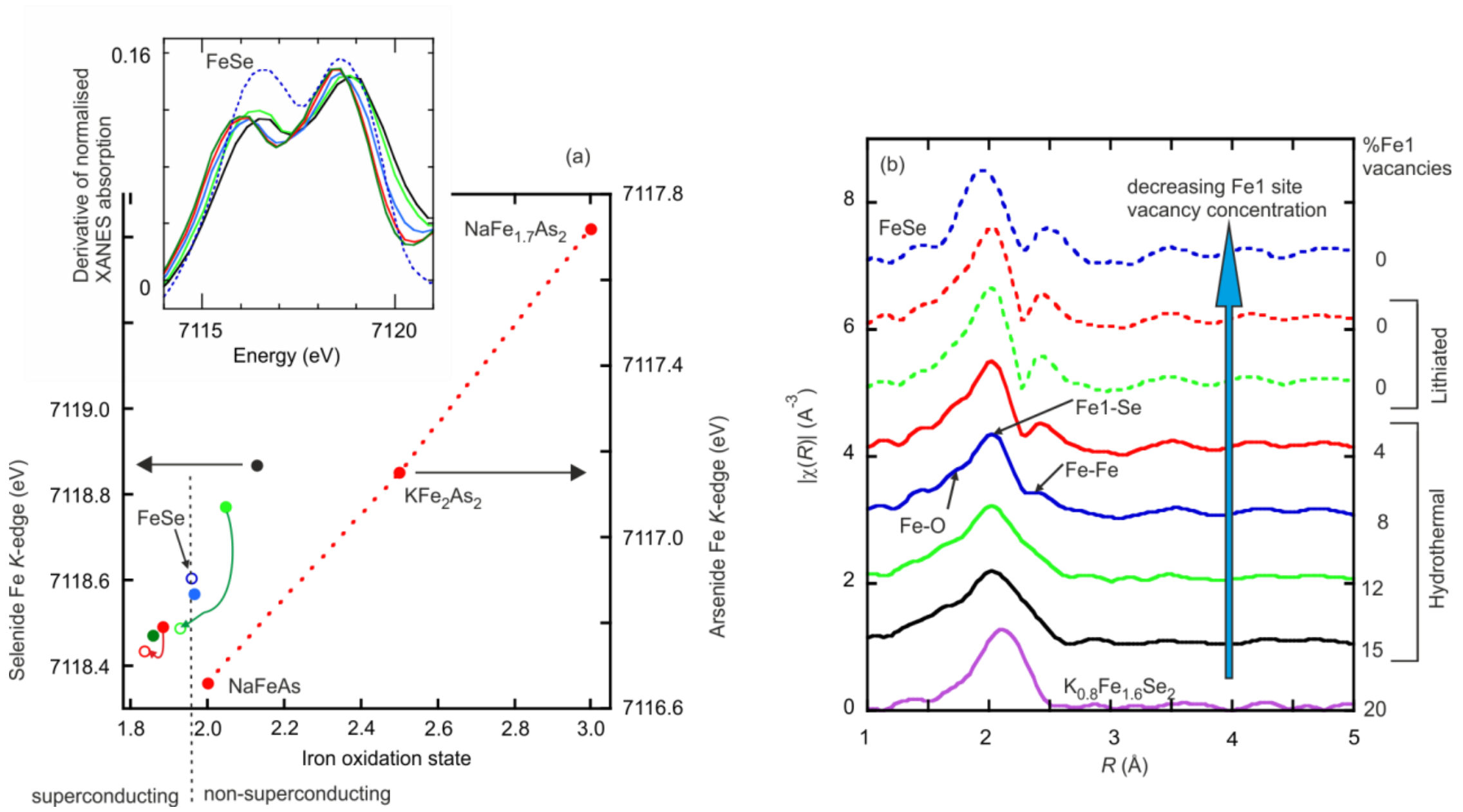

Figure 3. (a) Fe $K$-edge positions of hydrothermally-synthesised $Li_{1-x}Fe_x(OH)Fe_{1-y}Se$ samples (filled coloured symbols) plotted as a function of Fe oxidation state obtained from the refined compositions from diffraction data. The curved arrows show the evolution of the edge positions after lithiation (open coloured symbols). The inset shows the first derivative of the XANES absorption with the curves carrying the same colour as the points in the main figure. FeSe is included for comparison in both figures. The boundary between the superconducting and non-superconducting $Li_{1-x}Fe_x(OH)Fe_{1-y}Se$ samples is indicated. Iron arsenides[34] (red symbols) are included to calibrate the rate of change of edge position with oxidation state. (b) Comparison of the Fe $K$-edge EXAFS region for a series of hydrothermally-synthesised samples (coloured solid lines; similar colours for each sample are used in (a) and (b)) and lithiated samples (coloured dotted lines) compared with FeSe and $K_{0.8}Fe_{1.6}Se_2$ (fits are provided in Figures S8 & S9). The EXAFS region is extremely sensitive to the Fe content in the selenide layer (see Tables S3 & S4)

NPD data at 295K and 50K revealed no evidence for long range magnetic order. Superconducting samples with the highest Fe1 site occupancies showed a broad reflection at 5.565Å (Figure S2) which was invariant in intensity with temperature. It may arise from short range structural ordering of the Li and Fe ions in the hydroxide layers rather than magnetic order. Samples with larger cation vacancy concentrations in the selenide layer (up to 15%) did not exhibit these broad features (Figure S2) nor was there evidence in the SCXRD or NPD data for the long-range iron/vacancy order[10] found in the even more iron deficient (20% vacancies) "2-4-5" $A_{1-x}Fe_{2-2y}Se_2$ ($x \sim y \sim 0.2$) phases.

Semiquantitative Energy Dispersive analysis of X-rays (EDX) conducted using an FEI Quanta 650 FEG SEM equipped with an Oxford Instruments Aztec EDS detector produced Fe:Se ratios of about 1.1 : 1 with a 3–5 % uncertainty, consistent with the composition obtained from the crystallographic measurements and poorer in iron than the 1.5 : 1 ratio proposed in ref. 20 but in line with refs. 21 and 22. Exposure of the hydrothermally synthesised samples to air for one week broadened the superconducting transition, and reduced the shielding fraction (Figure S5), although the superconducting state was not completely destroyed.

Figure 2(c) shows that the Fe1 site occupancy controls whether the samples superconduct and the value of $T_c$. When this occupancy exceeds 95 %, the samples superconduct, and $T_c$ increases with increasing site occupancy. In the absence of significant compositional variation in other parts of the structure, a high iron occupancy in the selenide layer corresponds to a low Fe oxidation state. Computing the mean iron oxidation state from the refined composition for all the hydrothermally-synthesised samples probed by NPD and SCXRD shows that for iron oxidation states greater than +2 superconductivity is not observed, while reduction of iron leads to the appearance of superconductivity and $T_c$ increases as the formal oxidation state decreases (Figure 2(d)).

Preliminary ambient temperature $^{57}$Fe Mössbauer spectroscopy measurements on a superconducting sample (Figure S4) showed two paramagnetic doublets, both consistent with Fe(II). The isomer shift of the more intense doublet closely resembles that found in FeSe,[32] and the isomer shift of the minor component is consistent with high-spin $Fe^{2+}$ in the hydroxide layer.[33] Normal state magnetic susceptibility measurements produced a Curie-Weiss type dependence (Figure S5) consistent with a paramagnetic contribution from $S = 2$ moments carried by the $Fe^{2+}$ ions (tetrahedral $d^6$) on the $Li/_{Fe}$ site in the hydroxide layer. Exposure of samples to laboratory air for 1 week resulted in an increase in the Curie constant consistent with oxidation of these species to $Fe^{3+}$ (tetrahedral $d^5$), and also led to the partial destruction of superconductivity.

X-ray absorption spectroscopy at the Fe-$K$-edge was used as an additional probe of the Fe oxidation state and the structure. Analysis of the X-ray Absorption Near Edge Structure (XANES) region for hydrothermally synthesised samples, representative of the full range of $a$ lattice parameters probed by diffraction methods, produced edge positions spanning 0.34 eV, suggesting oxidation states spanning approximately 0.3 based on the behaviour of structurally related materials.[34] A plot of absolute edge position against the Fe oxidation state computed from diffraction measurements produced a linear dependence with a gradient similar to that found for related iron arsenides (Figure 3(a)).[34]

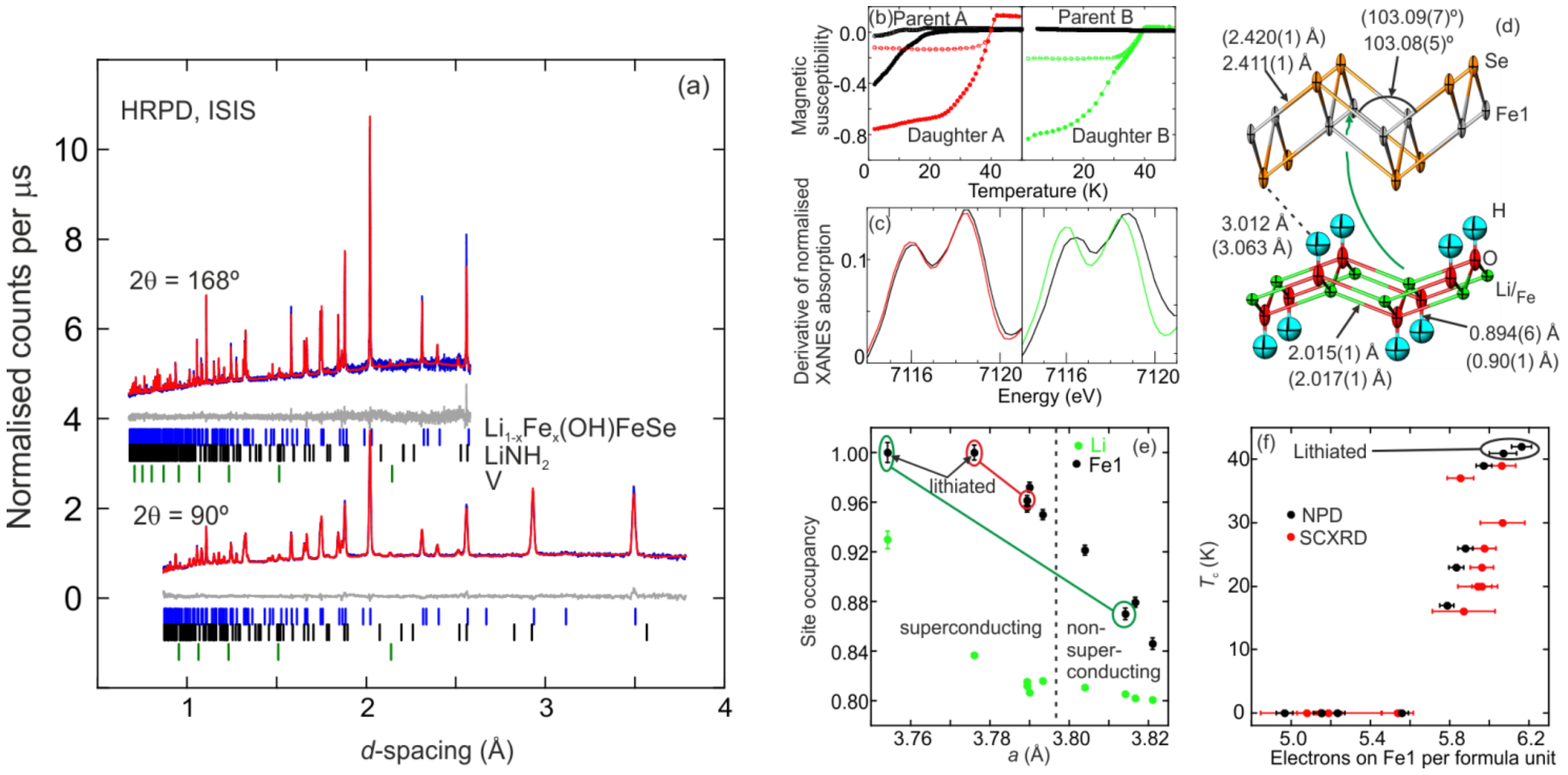

Figure 4. (a) Rietveld refinement against NPD data of the lithiated product daughter A with a refined composition $Li_{0.84}Fe_{0.16}(OH)FeSe$. The data from the 168° bank are displaced 4 units along the vertical axis. See also Table 1. (b) Enhancing $T_c$ or turning on superconductivity by lithiation of superconducting (Parent A) or non-superconducting (Parent B) hydrothermally synthesised materials. Daughter A was used for the μSR measurements (Figure 5(a)). (c) The reduction of Fe effected by lithiation as measured by the shifts in the Fe*K*-edge absorption energy. (d) Changes in refined structural parameters on lithiation of the superconducting Parent A to obtain Daughter A; 50% displacement ellipsoids are shown for the lithiated daughter product; refined parameters for Parent A are in parentheses. The arrow shows a possible pathway for migration of iron. (e) The correlation between Fe and Li site occupancies and the basal lattice parameter, *a*, for hydrothermally synthesised and lithiated samples (all results from NPD data). Parent and daughter samples are linked for clarity; in the lithiation of Parent A (red line) and Parent B (green line) to obtain the daughter products, the increase in the Fe1 site occupancy in the selenide layer is matched by the increasing Li occupancy in the hydroxide layer, so the freely-refined iron contents of parent and daughter samples do not vary by more than the uncertainty in the refined values (Table 1). (f) The correlation between superconducting $T_c$ and the average number of valence electrons per $Li_{1-x}Fe_x(OH)Fe_{1-y}Se$ formula unit assigned to the iron atoms in the selenide layers (a parameter that takes into account the iron deficiency in the selenide layer and the iron oxidation state).

Extended X-ray Absorption Fine Structure (EXAFS) spectra (Figure 3(b)) showed a sharp sample dependence which (Figures S8–S9) was determined by the vacancy concentration in the $Fe_{1-y}Se$ layers. The comparison of the EXAFS spectra at around $R$ = 2–3 Å for FeSe and the most Fe-poor $Li_{1-x}Fe_x(OH)Fe_{1-y}Se$ sample is similar to the comparison between FeSe and $K_{0.8}Fe_{1.6}Se_2$ (Figure 3(b) and Figure 2 in reference 35) and in related iron-deficient arsenides.[34] Refinement against the Fe *K*-edge EXAFS data (Figures S8–S9) produced ratios of iron on the Fe1 site in the selenide layer and on the $Li/_{Fe}$ site in the hydroxide layer consistent with the values obtained from diffraction, albeit with larger uncertainties (Table S4).

**Reductive lithiation.**

Analysis of the as-made hydrothermally synthesised samples shows that superconductivity is observed when the occupancy of the iron site (Fe1) in the selenide layers is high and iron is correspondingly reduced. A subsequent reductive lithiation step using lithium/ammonia solution was applied to two of the as-made samples (one non-superconducting and the other superconducting) which had been investigated by NPD. The crystal structure of the compounds was maintained, but with a significant increase in the interlayer cell parameter *c* and a decrease in the basal lattice parameter *a* (Table 1). When the sample of the non-superconducting hydroxide selenide with 13(1) % vacancies in the $Fe_{1-y}Se$ layers was lithiated the *c* lattice parameter increased by 4.3% and the basal *a* lattice parameter decreased by 1.6 %. In both cases the products of the post synthetic lithiation were superconductors with large volume fractions and $T_c$s exceeding 40 K (Figure 4(b)), higher than in any of the as-synthesised hydrothermal samples.

Rietveld analysis of NPD data (Figures 4(a) & S3, Table S1) from both lithiated samples revealed an increase in the occupancy of the Fe1 tetrahedral site in the selenide layers to 1.00(1) matched by a decrease in the Fe content of the $Li/_{Fe}$ shared site (Table 1, Figure 4(e)). The refinements constrained the $Li/_{Fe}$ site to be fully occupied so that the Li content of this site increased as its Fe content decreased. Overall Fe contents and the H occupancy were unconstrained in the refinements, but remained invariant under lithiation within the uncertainty. Figure 4(d) shows the shortest direct migration pathway for an Fe ion from the hydroxide layer moving to a site in the selenide layer 4.7 Å distant, presumably via the face of the $Se_4$ tetrahedron forming the target site. This migration may be enabled by the facts that the metal hydroxide layer is relatively flat and the SCXRD measurements show that the $Li/_{Fe}$ ellipsoid is elongated along *c* and may alternatively be modelled as a split site.[22] XANES measurements of the lithiated samples directly show the Fe K-edge shift arising from the reduction (Figures 3(a) and 4(c)), and EXAFS measurements (Figure 3(b)) show changes in the local structure consistent with the increased Fe content of the Fe1 site revealed by the NPD measurements. The lithiated samples were more air sensitive than the hydrothermally synthesized parents with only vestigial superconductivity evident after 1 week of air exposure (Figure S5).

Table 1. Changes in lattice parameters and refined site occupancies on lithiation from NPD data.

| | *a* (Å) | *c* (Å) | occ. Fe1 | occ. Li | Total Fe |
|---|---|---|---|---|---|
| Parent A | 3.7893(2) | 9.2617(6) | 0.961(4) | 0.812(2) | 1.15(1) |
| Daughter A | 3.7760(1) | 9.3512(2) | 1.004(5) | 0.837(2) | 1.165(5) |
| Parent B | 3.8142(3) | 9.1882(7) | 0.870(5) | 0.808(2) | 1.064(5) |
| Daughter B | 3.7542(1) | 9.5859(3) | 1.000(8) | 0.934(8) | 1.07(1) |

**Characterisation of the superconducting state.**

Muon-spin rotation (μSR) spectroscopy measurements on the lithiated sample $Li_{0.84}Fe_{0.16}(OH)FeSe$ ("Daughter A" in Table 1 and Figure 4(a)) are depicted in Figure 5(a). $B_{rms}$, the root-mean-square width of the magnetic field distribution experienced by the muon increases below $T_c$ due to the development of the superconducting vortex lattice and the behaviour of the average field $\langle B \rangle$ shows a diamagnetic response below $T_c$. These results confirm a superconducting volume fraction above 50 %. We extract an in-plane penetration depth, $\lambda_{ab}$, of 0.32(3) μm, where the relatively large error takes account of the uncertainty due to field-induced effects associated with the paramagnetic spins in the hydroxide layer. This places $Li_{0.84}Fe_{0.16}(OH)FeSe$ close to the main scaling line in a Uemura plot of $T_c$ against superfluid stiffness $\rho_s = c^2/\lambda_{ab}^2$ (inset to Figure 5(a)). Figure 5(b) shows the magnetisation as a function of applied magnetic field in a similar lithiated sample with $T_c$ = 40 K, as determined by SQUID magnetometry. This shows characteristics of a type-II superconductor. The lower critical field $H_{c1}$ is very small, so the Meissner effect is only apparent at lower temperatures, and a significant underlying paramagnetism presumably arises from the $Fe^{2+}$ moments in the hydroxide layer. The inset to Figure 5(b) shows the approximate evolution of $H_{c1}$ with temperature, as deduced from the susceptibility, calculated from the magnetisation data in the main figure. A correction for the effect of the paramagnetic $Fe^{2+}$ centres in the hydroxide layer (Figure S10) yields no evidence for the upper critical field $H_{c2}$, so we deduce that $\mu_0 H_{c2} > 7$ T, in line with the behaviour of other iron-based superconductors.

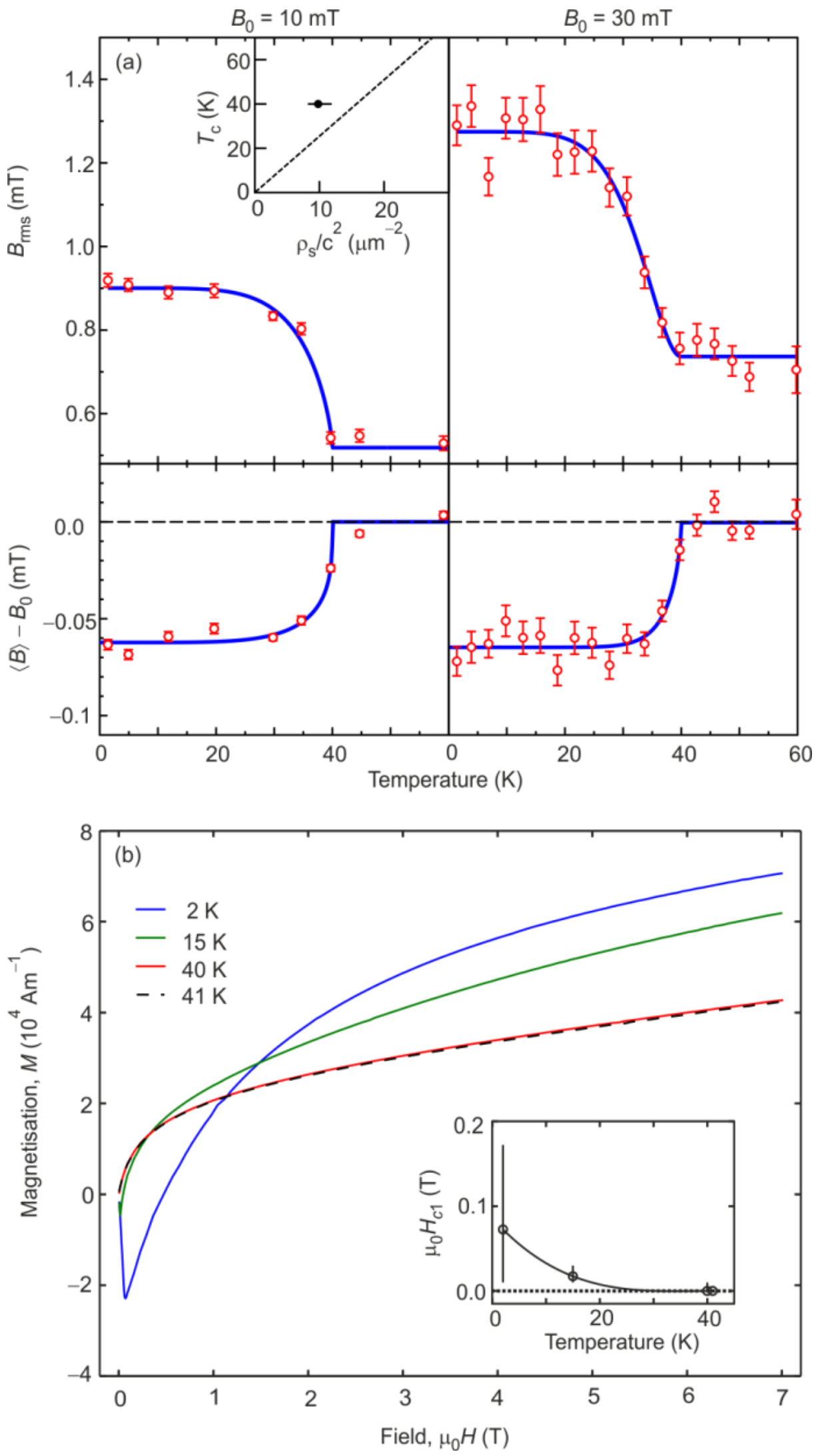

Figure 5. (a) The results of transverse-field muon-spin rotation spectroscopy on the lithiated sample "Daughter A" (see Figure 4; Table 1). While the diamagnetic response, measured by $\langle B \rangle - B_0$, which reflects the superconducting state only, is invariant with the applied transverse field, $B_0$, $B_{rms}$ increases with $B_0$ even in the normal state, which shows that there is a field-dependent contribution to the magnetic field distribution experienced by the muon that is likely due to the paramagnetic background originating from the $Fe^{2+}$ ions in the hydroxide layer. The different contributions to $B_{rms}$ act in quadrature ($B_{rms}^2 = \Sigma b_{rms}^2$). From the proportionality between the superconducting contribution to $B_{rms}$ and $1/\lambda_{ab}^2 = \rho_s/c^2$, where $\lambda_{ab} = 3^{1/4}\lambda$ is the in-plane penetration depth and $\rho_s$ is the superfluid stiffness, we extract $\lambda_{ab}$ = 0.32(3) μm. (b) Magnetisation as a function of magnetic field for a lithiated sample with $T_c$ = 40 K. In the inset, open circles illustrate the field at which the calculated susceptibility is equal to zero, bars illustrate the approximate width in H of the transition from the Meissner state to the vortex lattice state, and the dashed line is a guide to the eye.

**Conclusions.**

In conclusion we have demonstrated that hydrothermal synthesis under appropriate conditions yields $Li_{1-x}Fe_x(OH)Fe_{1-y}Se$ with $x \sim 0.2$ and with a highly variable $y$ that provides insight into the controlling parameters for superconductivity in iron selenides.

For $0.05 < y < 0.15$, the samples are non superconducting, but as the Fe deficiency, $y$, decreases and Fe is reduced, superconductivity emerges. Furthermore, superconductivity with the highest $T_c$s and shielding fractions can be turned on by reductive lithiation to intentionally reduce $y$ to zero: additional Li displaces some Fe ions from the hydroxide layer "reservoir" which migrate to completely fill the Fe site vacancies in the selenide layers, and the mean oxidation state of iron is reduced below +2. Figure 4(f) plots $T_c$ (defined to be 0 K for non-superconductors) against the number of 3*d* electrons associated with the iron atoms in the selenide layer per $Li_{1-x}Fe_x(OH)Fe_{1-y}Se$ formula unit, assuming the +2 oxidation state for Fe ions in the hydroxide layers. This quantity takes into account both the Fe1 site occupancy and the iron oxidation state. $T_c$ increases smoothly with increasing Fe electron count per formula unit once a threshold value is reached. These results provide a bridge between the two phases present in alkali metal iron selenide systems such as $K_{0.8}Fe_{1.6}Se_2$[9] where high Fe site occupancies and Fe oxidation states slightly below +2 are found in portions of the samples which show superconductivity, but the bulk of the sample is a magnetic insulator with a 20 % Fe deficiency and crystallographic ordering of the ensuing vacancies.[10] This underlines and quantifies the importance of structure and electron count in controlling superconductivity in iron selenide superconductors.

## ASSOCIATED CONTENT

### Supporting Information

Crystallographic data have been deposited at the FIZ, Karlsruhe with accession number 428281 (http://www.fiz-karlsruhe.de/icsd.html). These and additional characterization data are available free of charge via the Internet at http://pubs.acs.org.

## AUTHOR INFORMATION

### Corresponding Author

email: simon.clarke@chem.ox.ac.uk

### Author Contributions

HS and GMA prepared the samples, DNW, ALT, SJS, HS, and SJC (Clarke) performed the diffraction data collection and structural analysis. SJC (Cassidy) and SR collected and analysed the XAS data. FRF and SJB performed the μSR measurements, HS, GMA, BHW and SJC (Clarke) performed the magnetometry. PAB, SDF, SC and NM performed and analysed the Mössbauer spectroscopy. XL facilitated the work of HS in Oxford. SJC (Clarke) conceived the project and wrote the manuscript.

### Notes

The authors declare no competing financial interests.

## ACKNOWLEDGMENT

This work was funded by the UK Engineering and Physical Sciences Research Council (grant EP/I017844) and the Leverhulme Trust (grant RPG-2014-221). We also thank the Diamond Light Source Ltd. for the award of beamtime (MT9981 on I19; EE9697 on I11; SP11061 on B18) and studentship support for SJC (Cassidy), and the ISIS facility for the award of neutron beamtime including awards under the GEM Xpress programme. SJS acknowledges the support of a DFG Fellowship (SE2324/1-1). We are grateful for the assistance of beamline scientists at Diamond (G. Cibin, C. C. Tang) and ISIS (R. I. Smith, P. J. Baker), Dr N. Charnley, Oxford Earth Sciences, for assistance with SEM measurements and Dr R. Jacobs for assistance with TGA measurements.

**For Table of Contents Only.**

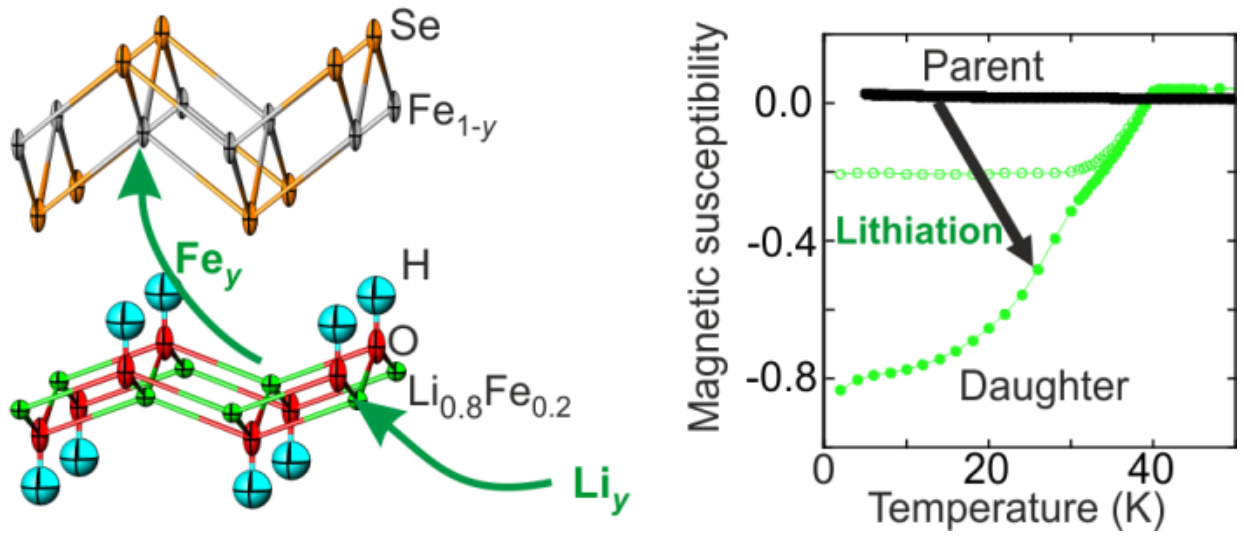

Chemical control of the iron content of the iron selenide layers of layered iron hydroxide selenides turns on high temperature superconductivity.